\pdfoutput=1
\documentclass[10pt,twocolumn,letterpaper]{article}

\usepackage[pagenumbers]{cvpr} 

\definecolor{cvprblue}{rgb}{0.21,0.49,0.74}
\usepackage[pagebackref,breaklinks,colorlinks,allcolors=cvprblue]{hyperref}

\usepackage{graphicx}
\usepackage{algorithm} 
\usepackage{algorithmic}  
\usepackage[algo2e,ruled,linesnumbered]{algorithm2e}
\SetKwInOut{Parameter}{parameter}
\usepackage{booktabs}
\usepackage{multirow}
\usepackage{svg}
\usepackage{cuted}

\newcommand{\customtitle}[2]{%
  \begin{center}
    {\large \textbf{#1}} \\[1.0em] 
    \vspace{1em} 
  \end{center}
}

\def\paperID{1223} 
\def\confName{CVPR}
\def\confYear{2025}

\begin{document}

\title{Progressive Rendering Distillation: Adapting Stable Diffusion for \\ Instant Text-to-Mesh Generation without 3D Data}

\author{
{
    Zhiyuan Ma$^{1,2}$ \enspace  
    Xinyue Liang$^{1,2}$ \enspace  
    Rongyuan Wu$^1$ \enspace  
    Xiangyu Zhu$^{3,4}$ \enspace  
    Zhen Lei$^{1,2,3,4}$\footnotemark[1] \enspace  
    Lei Zhang$^{1}$\footnotemark[1]
} \\
$^1$The Hong Kong Polytechnic University \\ 
$^2$Center for Artificial Intelligence and Robotics, HKISI CAS \\
$^3$State Key Laboratory of Multimodal Artificial Intelligence Systems, CASIA \\
$^4$School of Artificial Intelligence, University of Chinese Academy of Sciences, UCAS
}

\twocolumn[{
\maketitle
\begin{center}
    \captionsetup{type=figure}
    \includegraphics[width=1.0\textwidth]{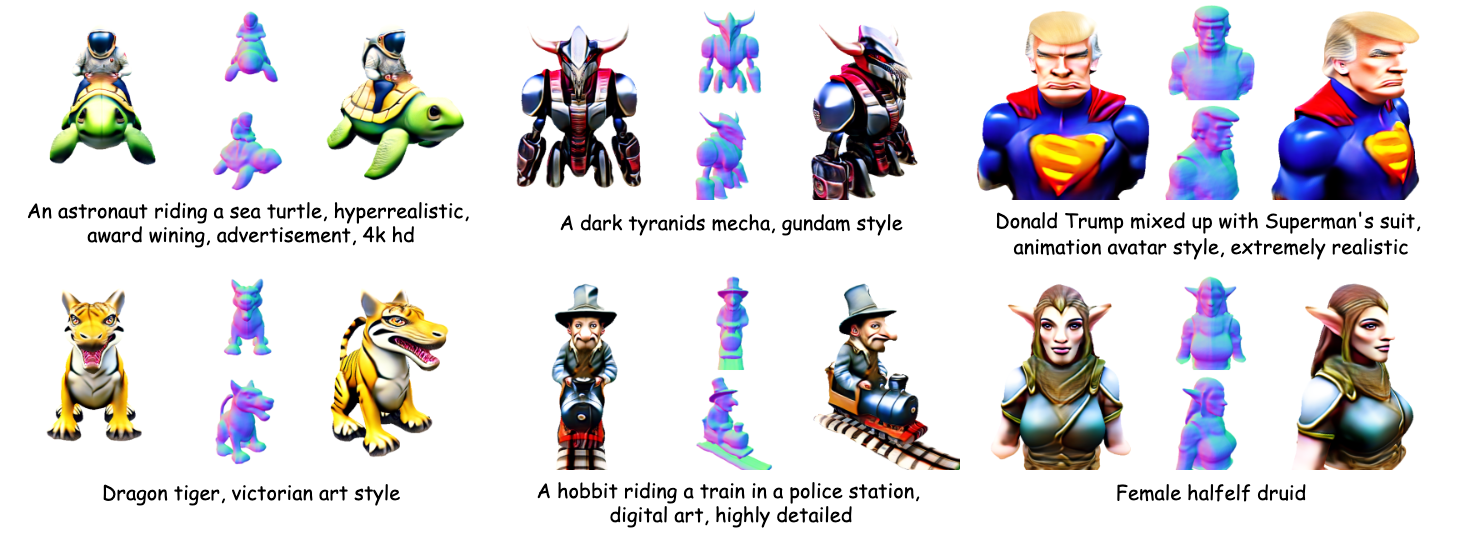}
    \captionof{figure}{Our method adapts Stable Diffusion~\cite{rombach2022high} to generate high-fidelity textured meshes in \textbf{1.2} seconds.}
    \label{fig:show_case}
\end{center}
}]

\footnotetext[1]{Corresponding authors.}

\begin{abstract}
    
It is highly desirable to obtain a model that can generate high-quality 3D meshes from text prompts in just seconds. While recent attempts have adapted pre-trained text-to-image diffusion models, such as Stable Diffusion (SD), into generators of 3D representations (e.g., Triplane), they often suffer from poor quality due to the lack of sufficient high-quality 3D training data. Aiming at overcoming the data shortage, we propose a novel training scheme, termed as Progressive Rendering Distillation (PRD), eliminating the need for 3D ground-truths by distilling multi-view diffusion models and adapting SD into a native 3D generator. In each iteration of training, PRD uses the U-Net to progressively denoise the latent from random noise for a few steps, and in each step it decodes the denoised latent into 3D output. Multi-view diffusion models, including MVDream and RichDreamer, are used in joint with SD to distill text-consistent textures and geometries into the 3D outputs through score distillation. Since PRD supports training without 3D ground-truths, we can easily scale up the training data and improve generation quality for challenging text prompts with creative concepts. Meanwhile, PRD can accelerate the inference speed of the generation model in just a few steps.  With PRD, we train a Triplane generator, namely TriplaneTurbo, which adds only $2.5\%$ trainable parameters to adapt SD for Triplane generation. TriplaneTurbo outperforms previous text-to-3D  generators in both efficiency and quality. Specifically, it can produce high-quality 3D meshes in 1.2 seconds and generalize well for challenging text input. The code is available at \href{https://github.com/theEricMa/TriplaneTurbo}{github.com/theEricMa/TriplaneTurbo}.
\end{abstract}

\section{Introduction}

Text-to-3D aims to create 3D content faithful to the given textual descriptions. 
The optimization-based text-to-3D methods can achieve high generation fidelity by using image diffusion priors~\cite{liu2023zero1to3,shi2023MVDream,long2023wonder3d,sweetdreamer,Ma2023GeoDream,qiu2023richdreamer,liu2023unidream,voleti2024sv3d} and score distillation techniques~\cite{poole2022dreamfusion,zhu2023hifa,sjc,wang2023prolificdreamer,katzir2023noisefree,yu2023textto3d,liang2023luciddreamer,ma2024scaledreamer} to optimize 3D representations~\cite{lin2023magic3d,Chen_2023_ICCV,tang2023dreamgaussian,hoellein2023text2room,liu2024headartist}. However, these approaches encounter bottlenecks in computational efficiency since they take hours to generate 3D textured meshes. Recent researches have shifted towards learning-based methods~\cite{lan2024ln3diff,gupta20233dgen,ntavelis2023autodecoding,huang2023textfield3d,cao2023large},  which generate 3D content through feedforward networks, reducing the latency to a few seconds per output. Unfortunately, existing 3D datasets~\cite{deitke2023objaversexl,downs2022google,collins2022abo,wu2023omniobject3d,pan2023aria} are much smaller compared to those used in training text-to-image generation models, while the 3D data therein suffer from texture quality and inconsistent object poses~\cite{liu2024direct}. Consequently, these approaches struggle to produce high-quality 3D outputs. As a promising alternative to the above mentioned methods, one can adapt pretrained text-to-image models, such as Stable Diffusion (SD) \cite{rombach2022high}, into generators of 3D representations~\cite{mercier2024hexagen3d,liu2024pi3d,li2024dual3d}. Recent studies have enabled the use of SD to generate 2D planes for 3D representations such as Triplane~\cite{chan2022efficient}. PI3D~\cite{liu2024pi3d} adapts SD to generate six 2D planes pre-optimized for 3D object reconstruction.  HexaGen~\cite{mercier2024hexagen3d} first trains a VAE to compress 3D objects into latent space, then adapts SD to generate the desired latents from text prompts. These methods leverage SD's prior knowledge for diverse text prompts while reducing training costs. However, their reliance on data-driven training limits the generalization performance.

In this paper, we propose to address the data shortage problem by using high-quality multi-view diffusion models~\cite{qiu2023richdreamer,shi2023MVDream,liu2023unidream} as teachers and performing  multi-view distillation to adapt SD into an instant text-to-mesh generator without using 3D ground-truth data. 
Data-free distillation techniques~\cite{lopes2017data,zhu2021data} have been applied in previous methods~\cite{ma2024scaledreamer,lorraine2023att3d,xie2024latte3d,li2024instant3d} to train native 3D generators from scratch, whereas they struggle to produce high-quality 3D outputs. We argue that if we can adapt SD for native 3D generation, instead of training from scratch, not only the training efficiency can be improved but also the generation quality can be enhanced. To our best knowledge, however, no such attempt has been made before. The primary obstacle lies in the training scheme of SD, which requires 3D  ground-truth data to supervise the denoising process. This requirement, unfortunately, conflicts with our goal to eliminate the dependency on 3D training data.

To address the above challenges, we propose \textbf{Progressive Rendering Distillation} (PRD), which enables 3D-data-free distillation. PRD achieves this goal by denoising the latent from random noise rather than 3D ground-truths. In each training iteration, PRD uses a few steps to progressively reverse noise to latent space using the adapted U-Net of SD. Then the denoised latent of each step is decoded to 3D content via the adapted VAE decoder. Multi-view teacher models are used to distill high-quality renderings through efficient score distillation techniques such as ASD~\cite{ma2024scaledreamer}.  An additional benefit of PRD is that by adapting the SD to generate 3D content in just four steps, the overall generation process can be much accelerated.
 
PRD is flexible in the choices of multi-view teacher models and the types of 3D representations. Unlike existing approaches using SD alone as the teacher~\cite{stable-dreamfusion,wang2023prolificdreamer,zhu2023hifa,GaussianDreamer,lin2023magic3d}, which risk the multi-view inconsistency and the Janus problem~\cite{armandpour2023re}, we employ multiple teachers in training.
Specifically, we employ  MVDream~\cite{shi2023MVDream} and SD to suppress the Janus problem and ensure text consistency in 3D content supervision.
Additionally, we use RichDreamer~\cite{qiu2023richdreamer} for geometry supervision through normal and depth.
With PRD, we adapt SD into a native 3D generator to produce geometry and texture Triplanes in 4 steps, which is named as \textbf{TriplaneTurbo} in the following context.
To enhance TriplaneTurbo's textured mesh quality, we employ SDF-based volumetric rendering~\cite{yariv2021volume} and mesh rasterization~\cite{wei2023neumanifold} for multi-view teacher supervision. 

In addition, to address the high GPU memory usage caused by multiple teachers and multi-view renderings in the training process, we introduce the \textbf{Parameter-Efficient Triplane Adapter} (PETA), which adds only $2.5\%$ trainable parameters to the frozen SD and effectively adapts it for 3D generation. Some results are shown in \cref{fig:show_case}. To our knowledge, \textit{this is the first parameter-efficient training method for direct 3D content generation from 2D diffusion models}, departing from full-parameter adaptation approaches. Our key contributions are summarized as follows:

\begin{itemize}
    \item We make the first attempt to adapt the pretrained 2D SD model into a native 3D generator without 3D data. With the proposed PRD scheme, we use multi-view diffusion models as teachers and distill SD into a four-step native 3D generator, namely TriplaneTurbo.
 
    \item  TriplaneTurbo adds only $2.5\%$ additional trainable parameters to the frozen SD for Triplane adapation. It marks the first use of parameter-efficient training to adapt 2D diffusion models for native 3D generation.
    
    \item TriplaneTurbo surpasses existing text-to-3D models not only in quality but also in speed, reducing text-to-mesh generation time to just 1.2 seconds. In addition, by scaling up the text training data, the model can generalize much better to complex text input.
    
\end{itemize}

\section{Related Work}


\noindent \textbf{Data-Driven Models for 3D Generation}. Employing 3D data to train generators has shown its effectiveness for single-category 3D generation such as  human faces~\cite{ma2023otavatar,chan2022efficient,an2023panohead,sun2023next3d,wu2023aniportraitgan,xiang2023gram,hong2022headnerf}, body shapes~\cite{saito2020pifuhd,liao2023high,xu2023efficient,hu2024structldm}, everyday objects~\cite{anciukevivcius2023renderdiffusion,muller2023diffrf,karnewar2023holodiffusion,liu2024vqa} , and structured room layouts~\cite{bahmani2023cc3d,wu2024blockfusion,meng2024lt3sd,feng2024prim2room,roessle2024l3dg,tang2024diffuscene}. This is mainly because for these specific categories, the training data are relatively easy to collect. However, text-to-3D generation requires open-category generation, bringing a rather different challenge. The difficulties in obtaining sufficient 3D training data largely limit the model generalizability across diverse text prompts~\cite{ma2024scaledreamer}, regardless of the type of used generation models (\eg, GAN~\cite{schwarz2020graf,niemeyer2021giraffe,goodfellow2020generative} or diffusion models~\cite{ho2020denoising,karras2022elucidating,song2020score}) or 3D representations (\eg, NeRF~\cite{mildenhall2021nerf,barron2021mip} or Gaussian Splatting~\cite{kerbl20233d,yu2024mip}). The publicly available 3D datasets~\cite{deitke2023objaversexl,downs2022google,collins2022abo,wu2023omniobject3d,pan2023aria} mostly contain only hundreds or thousands of samples, which are very hard, if not impossible, to train a generalized text-to-3D generation model. 

\begin{figure*}[!t] %
  \centering
  \includegraphics[width=1.0\textwidth]{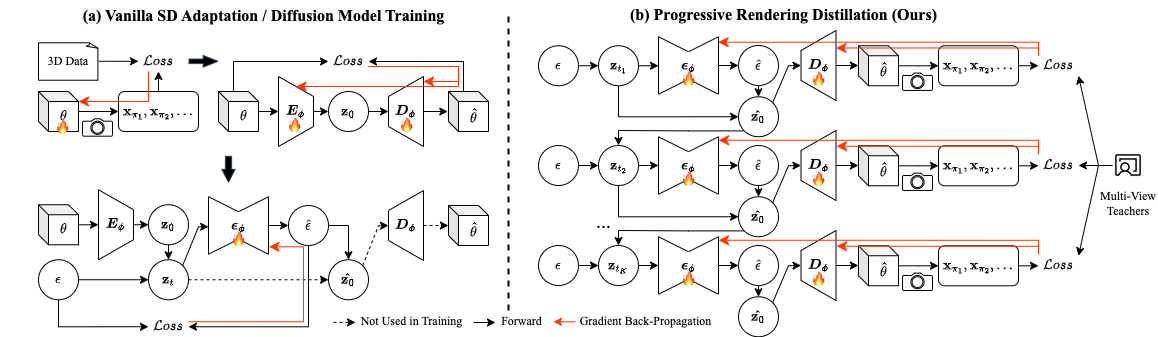}
  \caption{Comparison between (a) traditional SD adaptation and (b) our proposed progressive rendering distillation (PRD) for native 3D generation. Traditional approach requires ground-truth 3D representations $\theta$ and their latents $\boldsymbol{z}_0$ for each 3D sample to generate $\boldsymbol{z}_0$. Our proposed PRD scheme progressively denoises latents $\boldsymbol{z}_t$ initialized from random noise into $\boldsymbol{z}_0$, which are decoded to $\theta$, using multi-view diffusion models as teachers for distillation, eliminating the need for 3D data during adaptation and overcoming data scarcity.}
  \label{fig:prd}
\end{figure*}

\noindent \textbf{2D Diffusion Models for 3D Generation}. The difficulty of text-to-3D generation can be alleviated by leveraging 2D diffusion models as priors, thanks to their training on vast text-image pairs.  Early approaches like SDS~\cite{stable-dreamfusion} and VSD~\cite{cheng2023progressive3d} pioneered zero-shot text-to-3D generation by optimizing 3D representations (NeRF \cite{sjc,liu2023zero1to3,poole2022dreamfusion,lin2023magic3d}, mesh \cite{lin2023magic3d,wang2023prolificdreamer,chen2023fantasia3d,yang2024dreammesh}, 3DGS \cite{tang2023dreamgaussian,GaussianDreamer,GaussianDreamerPro,liang2023luciddreamer}) through score distillation \cite{poole2022dreamfusion,yu2023text,ma2024scaledreamer,liang2023luciddreamer,wang2023prolificdreamer,li2025connecting,wei2024adversarial,katzir2023noise,huang2024placiddreamer,wu2024consistent3d,zhou2024score,tang2023stable,zheng2025recdreamer,alldieck2024score,wang2023steindreamer,zhuo2024vividdreamer},
which serves as a bridge to transfer the generative capability of 2D diffusion models to 3D representations through rendered views. 
However, pre-trained text-to-image models like SD lack multi-view consistency, leading to the Janus problem~\cite{armandpour2023re}. MVDream \cite{shi2023MVDream} and subsequent works \cite{hu2024efficientdreamer,liu2025unidream} address this issue by camera-aware adaptations and synchronized multi-view generation, and they incorporate additional modalities such as normal~\cite{qiu2023richdreamer}, depth~\cite{fu2024geowizard,ke2024repurposing}, and CCM~\cite{liu2023syncdreamer,xu2025sparp} to enhance geometric quality. While achieving improved 3D generation results, these methods require computationally intensive score distillation~\cite{sargent2023zeronvs,Cascadezero123,shi2023MVDream,qiu2023richdreamer} or 3D reconstruction~\cite{hong2023lrm,gslrm2024,tang2024lgm,xu2024grm,wang2024crm,li2024craftsman,shi2023zero123plus,long2023wonder3d,liu2024mvboost}. Native 3D generators~\cite{wu2024direct3d,lan2024ln3diff,he2024gvgen,zhang2024clay,tang2023volumediffusion,hu2024structldm,xiang2024structured,deitke2023objaversexl,zhang2024clay,lin2025diffsplat,zhou2024diffgs,zhou2024diffgs,zhang2024gaussiancube,lan2024gaussiananything, ma2024diffspeaker} can reduce the generation time to seconds by directly producing 3D content without view rendering as proxies. For example, LN3Diff \cite{lan2024ln3diff} and GVGEN \cite{he2024gvgen} compress NeRF~\cite{mildenhall2021nerf} and 3DGS~\cite{kerbl20233d} into latent spaces using VAEs~\cite{kingma2013auto}, then train text-conditioned latent diffusion models~\cite{ho2020denoising}. However, these methods show limited generalizability across text prompts, as their performance is constrained by the insufficient text-3D training pairs. 

Some methods \cite{liu2024pi3d,mercier2024hexagen3d,shi2023MVDream,qiu2023richdreamer} leverage SD as their backbone to transfer text-to-image knowledge into text-to-3D generation. SD provides strong generative prior and improves the model generalizability~\cite{10735788,lin2025diffsplat,tai2024deffiller,zhang2024transparent,elizarov2024geometry,sun2023improving,wu2024one,sun2024pixel,wu2024seesr,yang2024pixel,chen2024adversarial}. Approaches like PI3D \cite{liu2024pi3d} adapt SD to generate multiple planes for constructing 3D space, yet their performance remains limited due to the insufficient 3D training data. Instead of adapting SD for multi-view generation, native 3D generation requires substantially more 3D data for effective adaptation. We propose to address the data insufficiency challenge by distilling knowledge from multi-view diffusion models into an SD-adapted native 3D generator, eliminating the need for 3D training data. While previous works like ATT3D~\cite{lorraine2023att3d} and ScaleDreamer \cite{ma2024scaledreamer} have investigated such a data-free training strategy, they employ multi-view distillation to train generators from scratch, and show limited performance due to the insufficient training scale. To overcome the huge training cost, we propose a cost-effective solution that combines multi-view distillation with SD-based native 3D generation. A fundamental challenge to achieve this goal is how to adapt SD for 3D generation without ground-truth data. We address this by proposing a novel Progressive Rendering Distillation scheme, which not only eliminates the need of 3D ground-truths but also enables few-step generation. 

\section{Method}

\subsection{Preliminary}

\textbf{Stable Diffusion (SD)} performs diffusion in latent space for efficient text-to-image generation. Its VAE encoder $\emph{E}_{\phi^{\mathrm{SD}}}$ compresses an image $\boldsymbol{x}$ into a latent code $\boldsymbol{z}$, while its decoder $\emph{D}_{\phi^{\mathrm{SD}}}$ reconstructs the image. Given text prompt $y$, a U-net $\boldsymbol{\epsilon}_{\phi^{\mathrm{SD}}}$ predicts noise $\boldsymbol{\epsilon}$, which is added to $\boldsymbol{z}_t = \alpha_t \boldsymbol{z} + \sigma_t \boldsymbol{\epsilon}$, where timestep $t \in {1, \dots, T}$ controls noise level via scalars $\alpha_t$ and $\sigma_t$. Generation proceeds by iteratively denoising from $\boldsymbol{z}_T$ to prompt-aligned $\boldsymbol{z}_0$. At each step, the U-net estimates noise $\hat{\boldsymbol{\epsilon}} = \boldsymbol{\epsilon}_{\phi^{\mathrm{SD}}}(\boldsymbol{z}_t; t,y)$ to compute $\hat{\boldsymbol{z}_0} = \frac{\boldsymbol{z}_t - \sigma_t \hat{\boldsymbol{\epsilon}}}{\alpha_t}$, denoted as $\hat{\boldsymbol{z}_0} = \boldsymbol{z}_{\phi^{\mathrm{SD}}}(\boldsymbol{z}_t; t, y)$. Results can be refined through additional diffusion steps $t' < t$. The final latent is decoded to an image via $\hat{\boldsymbol{x}} = \emph{D}_{\phi^{\mathrm{SD}}}(\hat{\boldsymbol{z}_0})$.

\noindent \textbf{Score Distillation}. 2D diffusion models can optimize 3D representations $\theta$ through differentiable rendering $\boldsymbol{x}_{\pi} = \boldsymbol{g}(\theta, \pi)$~\cite{kerbl3Dgaussians,son2024dmesh,shen2021deep}, which produces images $\boldsymbol{x}_{\pi}$ from camera view $\pi$. Here 2D diffusion models serve as a metric $\mathcal{L}(\boldsymbol{x}_{\pi}; \pi, y)$ that measures the consistency between $\boldsymbol{x}_{\pi}$ and the text conditions $y$. The 3D representation is optimized using gradient $\nabla_\theta\mathcal{L}(\boldsymbol{x}_{\pi}; \pi, y) = \nabla{\boldsymbol{x}_{\pi}}\mathcal{L}(\boldsymbol{x}_{{\pi}}; \pi, y)\frac{\partial \boldsymbol{x}_{\pi}}{\partial \theta}$, which also trains the native 3D generator. The computation of $\nabla{\boldsymbol{x}_{\pi}}\mathcal{L}$ depends on the chosen score distillation method.

\subsection{Progressive Rendering Distillation}
\label{sec:prd}

We now detail our proposed training scheme for adapting SD as a native 3D generator. Traditional adaptation approaches require preparing ground-truth 3D representations $\theta$ and their corresponding latents $\boldsymbol{z}$ for each 3D sample in the dataset. The U-net is adapted to denoise diffused latents $\boldsymbol{z}_t$ by minimizing the noise prediction mean squared error (MSE). \cref{fig:prd}(a) illustrates this paradigm, which has been used by several native 3D generators \cite{liu2024pi3d,wang2023rodin,zhang2024clay,wu2024direct3d}. However, this paradigm faces limitations in both the quantity and quality of available 3D representations $\theta$, as existing 3D datasets lack sufficient high-quality data for training text-to-3D generators. Actually, the pretrained SD models already possess denoising capabilities for image generation. In other words, pretrained SD is well-trained for a Markov Chain to reverse $\boldsymbol{z}_T = \epsilon \sim \mathcal{N}(0, \boldsymbol{I})$ to $\boldsymbol{z}_0$ with its U-net $\boldsymbol{z}_{\phi^{\mathrm{SD}}}$, and decode $\boldsymbol{z}_0$ to image with its decoder $\emph{D}_{\phi^\mathrm{SD}}$. Our goal is to modify the Markov chain by transforming $\boldsymbol{z}_{\phi^{\mathrm{SD}}}$ and $\emph{D}_{\phi^\mathrm{SD}}$ into 3D generators $\boldsymbol{z}_{\phi^{\mathrm{3D}}}$ and $\emph{D}_{\phi^\mathrm{3D}}$, from which the 3D representations $\theta$ can be decoded. Note that our modification of the Markov chain differs from the traditional diffusion model adaptation objectives, as it requires neither ground-truth latents $\boldsymbol{z}_0$ nor their noise-diffused variants $\boldsymbol{z}_t$ in the training process. 

Specifically, at the beginning of the Markov chain, the network takes random noise $\boldsymbol{\epsilon}$ as input to represent $\boldsymbol{z}_T$. At each step, the current state $\boldsymbol{z}_t$ is used to estimate $\hat{\boldsymbol{z}_0}$ through $\boldsymbol{z}_{\phi^{\mathrm{SD}}}$, which is then decoded to 3D output $\hat{\theta}$ with $\emph{D}_{\phi^\mathrm{SD}}$. The 3D output $\hat{\theta}$ is used to render images $\boldsymbol{x}_{\pi_1}, \boldsymbol{x}_{\pi_2}, \dots $ at camera views $\pi_1, \pi_2, \dots$ and receive supervision from multi-view teachers via score distillation, while the estimated $\hat{\boldsymbol{z}_0}$ is diffused to the next timestep $t'$ as $\boldsymbol{z}_{t'} = \alpha_{t'}\hat{\boldsymbol{z}_0} + \sigma_{t'}\boldsymbol{\epsilon}$ for subsequent operations. We name this training scheme Progressive Rendering Distillation (PRD), as illustrated in \cref{fig:prd}(b). From the total $T$ timesteps, we select a decreasing sequence of $K$ timesteps $T = t_1 > t_2 > \cdots > t_K = T/K$ with uniform spacing to perform score distillation from multi-view teachers. The gradient at each step is not backpropagated to previous steps; therefore, we can largely reduce the GPU memory usage and prevent gradient explosion \cite{clark2023directly,prabhudesai2023aligning,wu2024deep,xu2024imagereward}. Since $\phi^{\mathrm{3D}}$ is initialized from $\phi^{\mathrm{SD}}$, this specialized gradient detachment strategy still maintains good convergence. 
The pseudo code of our algorithm is provided in \cref{alg:progressive rendering distillation}.
While the time cost of our training strategy increases with the increase of $K$, we can ensure the model generates high-quality results in just a few steps, thereby accelerating inference. We set $K=4$ to balance quality and speed. 

\begin{figure*}[!t] 
  \centering
  \includegraphics[width=1.0\textwidth]{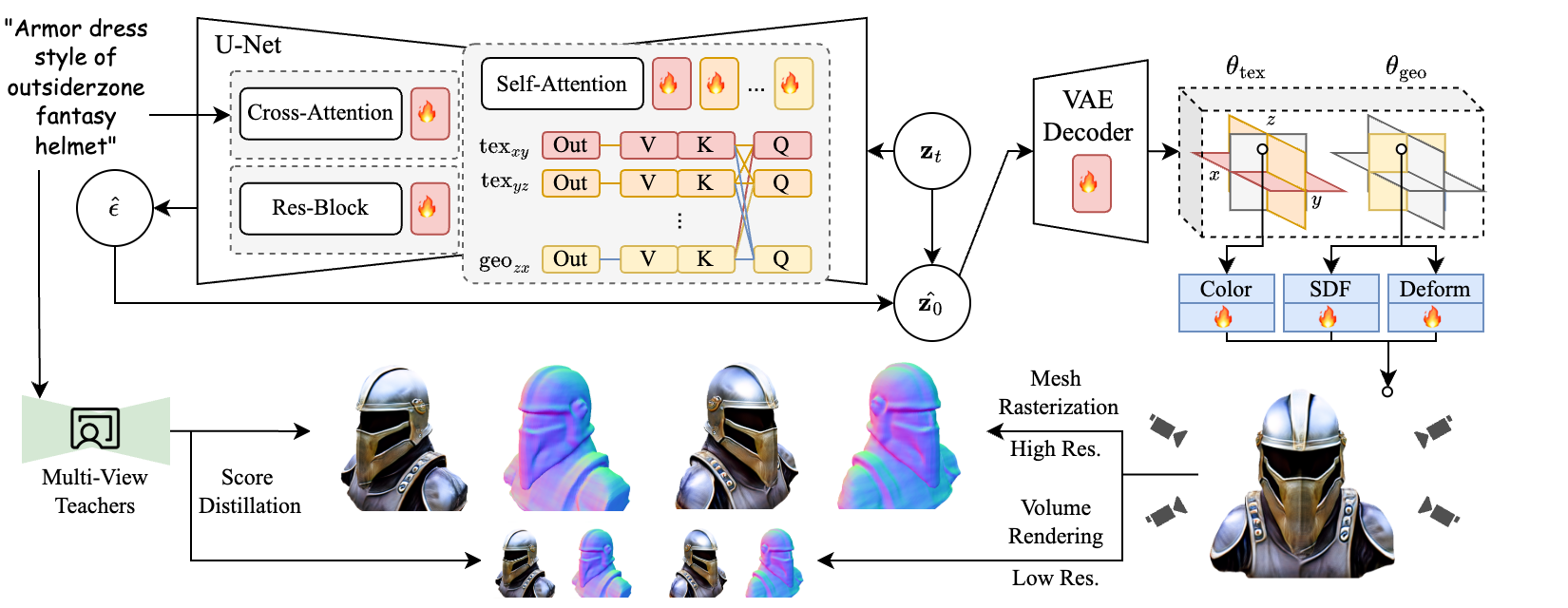}
  \caption{Illustration of TriplaneTurbo: an SD-adapted native 3D generator using our PRD scheme. Our model generates six feature planes comprising geometry Triplane $\theta_{\mathrm{geo}}$ and texture Triplane $\theta_{\mathrm{tex}}$ in 4 steps. We introduce Parameter Efficient Triplane Adaptation (PETA), which requires only $2.5\%$ additional parameters for adaptation. The parameter arrangement is illustrated in the figure.}
  \label{fig:peta}
\end{figure*}

\subsection{Parameter-Efficient Triplane Adaptation}

While various 3D representations could be employed by our PRD scheme, in this paper we demonstrate an exemplar solution using Triplanes as the representation. We denote our adapted model as \textbf{TriplaneTurbo} with parameters $\phi^{\mathrm{3D}}$. Specifically, TriplaneTurbo adapts SD to generate a 3D representation consisting of two Triplanes~\cite{chan2022efficient} $\theta = (\theta_{\mathrm{geo}}, \theta_{\mathrm{tex}})$: a geometry Triplane $\theta_{\mathrm{geo}}$ storing Signed Distance Function (SDF) and  deformation values for mesh extraction,  and a texture Triplane $\theta_{\mathrm{tex}}$ containing RGB attributes for painting texture on the mesh. For each point in 3D space, we use a two-layer MLP to decode its SDF value. The same process applies to texture and mesh deformation \cite{shen2021deep,wei2023neumanifold}. This separation of geometry and texture planes follows the work in~\cite{gao2022get3d,wu2024tpa3d,sun2022ide}.
We set the Triplane resolution as $256\times256$, and replace the last convolution in SD's decoder to output $32$ channels. The Triplanes' output has a dimension of $6\times256\times256\times32$ in feature space. Due to the $8\times$ compression of the VAE, this corresponds to $6\times32\times32\times4$ in latent space, distinct from the $1\times64\times64\times4$ latent generated by pretrained SD. To enable interaction between the six feature planes, we follow existing approaches \cite{long2023wonder3d,shi2023MVDream,wang2023imagedream,shi2023zero123plus} to adapt the U-net's self-attention \cite{rombach2022high,vaswani2017attention} to allow cross-plane attention. Unlike existing works that fully retrain SD \cite{Cascadezero123,shi2023MVDream,mercier2024hexagen3d,liu2024pi3d,li2024dual3d}, which can lead to catastrophic forgetting \cite{shi2023MVDream,liu2024pi3d}, we propose a parameter-efficient adaptation approach.
The core of our design lies in the fact that each of the six feature planes maintains its own unique feature distribution. Therefore, plane-specific characteristics must be incorporated into the adaptation process. As illustrated in \cref{fig:peta}, our adaptation modifies the convolution, self-attention, and cross-attention layers. We name our approach \textbf{Parameter-Efficient Triplane Adaptation (PETA)}.

As shown in \cref{fig:peta}, for the convolution blocks (Res-Blocks) and cross-attention layers, we implement LoRA~\cite{yeh2023navigating,hu2021lora} for parameter-efficient adaptation, and process the six planes uniformly. The plane-specific adaptations are then applied to the self-attention. For the self-attention blocks, we apply distinct LoRA layers~\cite{hu2021lora} to the $\mathrm{to\_Q}$, $\mathrm{to\_K}$, $\mathrm{to\_V}$, and $\mathrm{to\_Out}$ linear layers when processing each of the six feature planes. 
In each linear transformation within a self-attention block, the linear projection with multiple LoRAs is implemented in two steps. First, the frozen linear layer batch processes the features extracted from all the six planes together. Then, separate LoRA transformations are applied independently to the features of each plane.
This adaptation maintains low computational overhead while effectively introducing plane-specific processing during attention calculations across the six feature planes. 
It can be applied with other techniques like AdaLoRa~\cite{zhang2023adalora} and Vera \cite{kopiczko2023vera}. We leave this for further exploration. We set the LoRA rank to $16$ by default. While this adaptation adds only $2.5\%$ of the parameters to the SD model, it effectively enables native 3D generation.

\subsection{Distillation Details}
\label{sec:distillation_detail}

Since PRD eliminates the need for 3D data by referring to multi-view teachers for distillation, using \textbf{multiple teachers} allows us to combine their strengths while mitigating individual biases. 
Most previous works~\cite{poole2022dreamfusion,chung2023luciddreamer,lin2023magic3d,lorraine2023att3d,tang2023dreamgaussian} use SD model (parametrized by $\phi^{\mathrm{SD}}$) as the teacher for its ability to generate high-fidelity, text-consistent images. However, SD lacks camera-awareness, which can lead to the Janus problem~\cite{armandpour2023re}. MVDream \cite{shi2023MVDream} (MV, parametrized by $\phi^{\mathrm{MV}}$) addresses this by generating four camera-conditioned views simultaneously, but at the cost of reduced prompt consistency~\cite{wang2023imagedream}. 
While SD and MV complement each other, both of them focus on RGB rendering and provide no direct supervision on geometry. We further incorporate RichDreamer \cite{qiu2023richdreamer} (RD, parameterized by $\phi^{\mathrm{RD}}$), a model that generates four-view normal and depth maps based on text prompts. The score distillation guidance $\mathcal{L}\left(\boldsymbol{x}_\pi ; \pi, y\right)$ in our implementation thus integrates SD, MV, and RD.
At each PRD step (see \cref{sec:prd} and \cref{fig:prd}), given a text prompt $y$ and generated 3D representation $\hat{\theta}$ from $\boldsymbol{z}_{\phi^{\mathrm{3D}}}$ and $\emph{D}_{\phi^{\mathrm{3D}}}$, we sample four views $\pi_1, \dots, \pi_4$ at uniform azimuth intervals. We sample one view $\pi$ from ${\pi_1, \dots, \pi_4}$ to compute $\mathcal{L}_{\mathrm{SD}} = \mathcal{L}_{\phi^{\mathrm{SD}}}(\boldsymbol{x}_{\pi}; \pi, y)$. For MV, all the four views are used to compute $\mathcal{L}_{\mathrm{MV}} = 
\mathcal{L}_{\phi^{\mathrm{MV}}}(\boldsymbol{x}_{\pi_1},\dots,\boldsymbol{x}_{\pi_4};\pi_1,\dots,\pi_4,y)$. RD operates on concatenated normal and depth renderings $\boldsymbol{x}^{'}_{\pi_1},\dots,\boldsymbol{x}^{'}_{\pi_4}$, and its objective is $\mathcal{L}_{\mathrm{RD}} = \mathcal{L}_{\phi^{RD}}(\boldsymbol{x}^{'}_{\pi_1},\dots,\boldsymbol{x}^{'}_{\pi_4};\pi_1,\dots,\pi_4,y)$. Among existing score distillation methods~\cite{poole2022dreamfusion,yu2023text,ma2024scaledreamer,liang2023luciddreamer,wang2023prolificdreamer,li2025connecting,wei2024adversarial,katzir2023noise,huang2024placiddreamer,wu2024consistent3d}, we adopt Asynchronous Score Distillation (ASD) \cite{ma2024scaledreamer} for its pioneered efficiency in training deep text-to-3D generators. For 3D rendering, as illustrated in \cref{fig:peta}, we combine volumetric rendering~\cite{yariv2021volume}  with mesh rasterization~\cite{wei2023neumanifold}  to overcome the instability of pure mesh supervision~\cite{xu2024instantmesh}. See \cref{sec:more_implementation} for more details.

\begin{figure*}[!t]
    \centering
    \includegraphics[width=\textwidth]{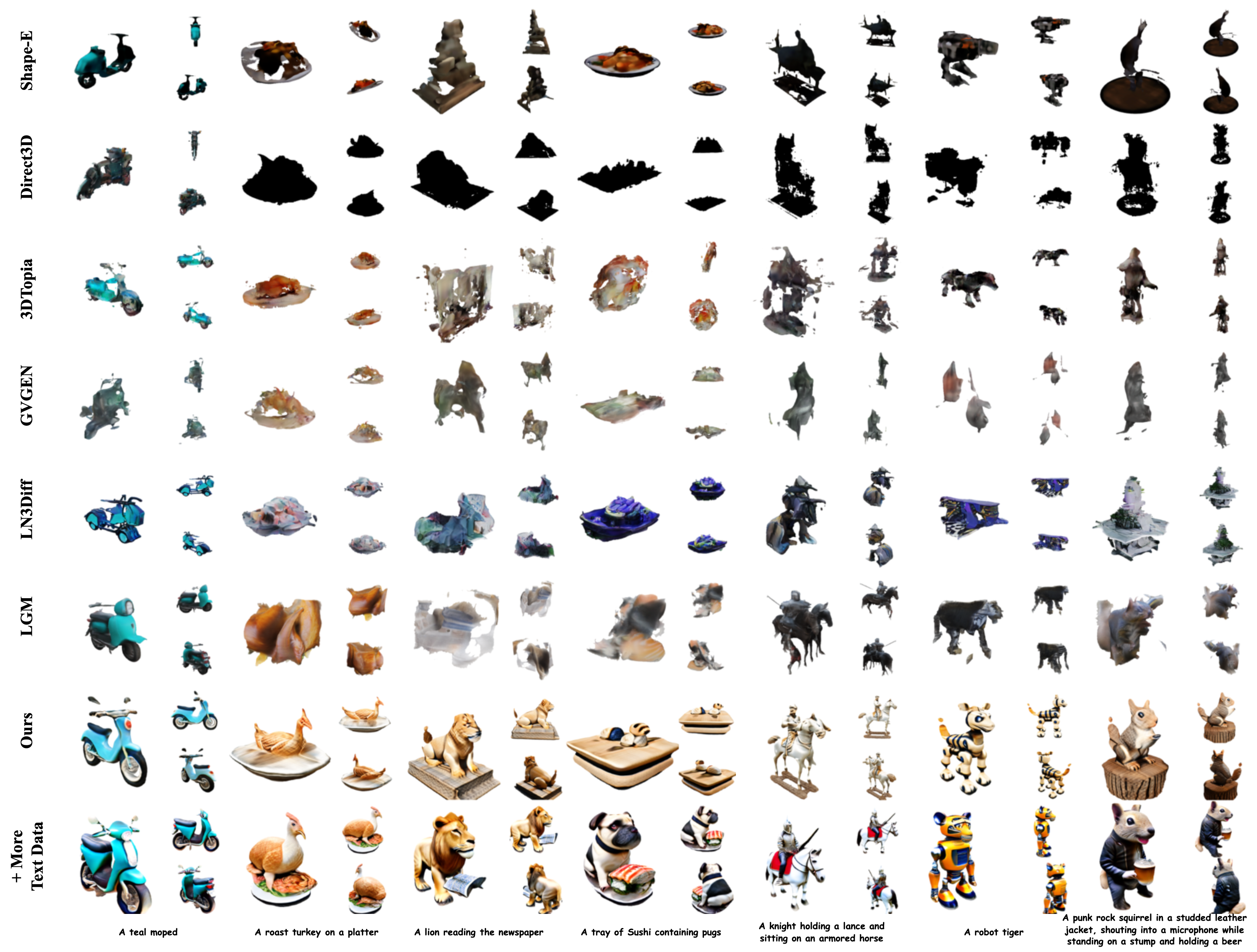}
    \vspace{-3mm}
    \caption{Qualitative comparison of text-to-mesh generation results by competing methods. Please refer to \cref{sec:comparison} for details.
    }
    \label{fig:comparison}
\end{figure*}

\section{Experiments}

\subsection{Experimental Settings}

\textbf{Implementation Details}. For a fair comparison with existing methods, we use captions of the 3D objects~\cite{deitke2023objaversexl} provided by~\cite{hong20243dtopia} to train our model, which comprises a total of 360K text prompts. We employ SD v2.1-base \cite{sdv2.1base} as the base model. Our model is trained for 15K iterations with a learning rate of 2e-4, costing 40 hours on 8 NVIDIA H20 GPUs. We further collect 1.6 million text prompts to evaluate the benefit of data scaling supported by our method; the detail is provided in \cref{sec:more_corpus}. Additional training details and loss weights are provided in \cref{sec:more_implementation}.

\textbf{Compared Methods}. 
Our proposed PRD aims for fast text-to-mesh generation. Therefore, we compare it against state-of-the-art approaches that can generate textured meshes within one minute, including Shape-E \cite{jun2023shape}, Direct3D~\cite{liu2024direct}, 3DTopia \cite{hong20243dtopia}, GVGEN \cite{he2024gvgen}, LN3Diff \cite{lan2024ln3diff}, LGM \cite{tang2024lgm} and PI3D \cite{liu2024pi3d}. 
Note that we do not compare with some relevant  works~\cite{mercier2024hexagen3d,xie2024latte3d} since their codes/models are not publicly available and they employ different evaluation protocols. 
In our experiments, the results of 
Shape-E, Direct3D, 3DTopia, GVGEN, LN3Diff and LGM
are obtained by running their publicly available models. 
For PI3D, we can only perform quantitative comparisons with it by copying its results from the original paper, but we cannot perform visual comparisons with it since its code/model is not yet publicly available.

\textbf{Evaluation Protocol}. 
We employ the protocol used in our competing methods~\cite{liu2024pi3d,tang2024lgm,lan2024ln3diff,he2024gvgen,hong20243dtopia,jun2023shape} to evaluate our PRD model. Specifically, we use the ViT-B/34 CLIP model~\cite{radford2021learning} to evaluate the test prompts from the DreamFusion gallery~\cite{poole2022dreamfusion}. As in~\cite{liu2024pi3d}, we render the generated 3D results at 512 resolution from four viewpoints at 15° elevation across four azimuth angles: 0°, 90°, 180° and 270°. Under these views, we evaluate the performance of competing methods using CLIP Score~\cite{hessel2021clipscore} (C.S.) and CLIP R-precision (R@1). To ensure a fair comparison for instant text-to-mesh generation, we exclude the post-processing steps such as SDS refinement~\cite{poole2022dreamfusion}, as they will prolong the generation time by several minutes. For models~\cite{tang2024lgm,he2024gvgen} using Gaussian Splatting representations, we use the conversion script from~\cite{tang2024lgm} to generate textured meshes.

\subsection{Experimental Results}
\label{sec:comparison}

We showcase qualitative comparisons of the competing methods in \cref{fig:comparison} using challenging test prompts. 
One can see that most of the existing methods struggle to generate satisfactory outputs. Our method, in comparison, produces better quality results with complete and vivid meshes. The quantitative results are reported in \cref{tab:comparision}. Again, our method demonstrates superior performance to its competitors in all the metrics. 
Existing methods fall short in the consistency of object poses and in the capability of handling complex prompts. Our PRD method addresses these challenges, enabling faster inference speed and improving the model performance by data scaling.



\textbf{Handling Inconsistent Generation Poses}. As shown in \cref{fig:comparison}, existing models often generate objects with incorrect poses. For prompts like \textit{`A teal moped'}, the competing methods like Direct3D~\cite{wu2024direct3d}, Shape-E~\cite{jun2023shape} and GVGEN~\cite{he2024gvgen} generate objects with misaligned orientations, such as facing sideways or backwards. This issue stems from the inconsistent object orientations in 3D training datasets, which are difficult to detect and correct automatically. Current methods struggle with noisy training data, leading to pose ambiguity and geometric defects. Our approach addresses this challenge by learning from multi-view teachers. Though our teachers are trained on noisy 3D datasets and may occasionally provide incorrect directional guidance, our PRD scheme exposes the 3D outputs to multi-view teachers $K$ times per iteration (see \cref{sec:prd}). Through the  iterative distillation process, we substantially decrease the impact of incorrect directional guidance, ensuring consistent pose alignment in the 3D outputs.

\textbf{Handling Complex Text Prompts}. The failure of our competing methods mainly stems from their reliance on the existing text-3D paired datasets, which are not comprehensive enough compared to the diverse user inputs. Therefore, trained on these data, existing methods fail to produce good results when the input is complex. For example, existing methods might perform well when the input prompt is \textit{`A robot'} but fail when the prompt becomes \textit{`A robot tiger'}. 
The quality of existing 3D datasets is also compromised by the absence of creative concepts. Creating 3D models for imaginative concepts like \textit{`A robot tiger'}
demands significant time and expertise from 3D artists. As a result, existing 3D datasets are largely limited to common everyday objects. Models trained on these limited datasets fail to generalize effectively to imaginative or complex prompts. We solve this problem by adapting SD as a 3D generator and inheriting its generative power, and more crucially, by introducing a training scheme that completely eliminates the need of 3D data. Our training scheme not only improves the quality of results on the existing training corpus, as shown by the qualitative and quantitative results in \cref{fig:comparison} and \cref{tab:comparision}, but also enhances the capability to handle complex prompts by expanding the available training data, as detailed below.

\begin{table}[!t]
\setlength{\belowcaptionskip}{-4.mm}
\setlength{\abovecaptionskip}{1.5mm}
\centering
\begin{tabular}{@{}lll|l@{}}
\toprule
         & C.S. $\uparrow$ & R@1$\uparrow$ & Latency (/sec) \\ \midrule
Shape-E~\cite{jun2023shape}  & 55.1               &  27.1           & 13.0       \\
Direct3D     & 60.8               & 4.33             & 16.0         \\ 
3DTopia~\cite{hong20243dtopia}  & 59.7                & 11.2            & 23.7      \\
PI3D$^{*}$~\cite{liu2024pi3d}     &  65.9              &  25.2           & 3.00       \\ 
GVGEN~\cite{he2024gvgen}     & 51.1             & 2.44           & 49.2      \\
LN3Diff~\cite{lan2024ln3diff}  & 55.9               & 5.09            & 8.16      \\
LGM~\cite{tang2024lgm}     & 67.4               & 28.3            & 56.1      \\ \midrule
Ours     &  68.2          & 32.3          & \textbf{1.23}      \\ 
+More Text Data     &  \textbf{75.1}          & \textbf{46.0}          & \textbf{1.23}      \\ \bottomrule
\end{tabular}
\caption{Quality and speed comparison for text-to-mesh generation. $^{*}$indicates that the values are quoted from the original papers. }
\label{tab:comparision}
\end{table}

\textbf{Fast Inference Speed}. Our PRD method also demonstrates superior computational efficiency. As shown in \cref{tab:comparision}, we evaluate the average inference latency from text input to textured mesh output across all test prompts on the H20 GPU device. While some methods~\cite{tang2024lgm,he2024gvgen} require dozens of steps, our PRD approach enables the native 3D generator to produce quality results in just $K$ steps. With the suggested setting of $K=4$, as shown in \cref{tab:comparision}, our model achieves sub-second latency for text-to-mesh generation, significantly outperforming previous methods.

\begin{figure*}[!t]
    \centering
    \includegraphics[width=0.93\linewidth]{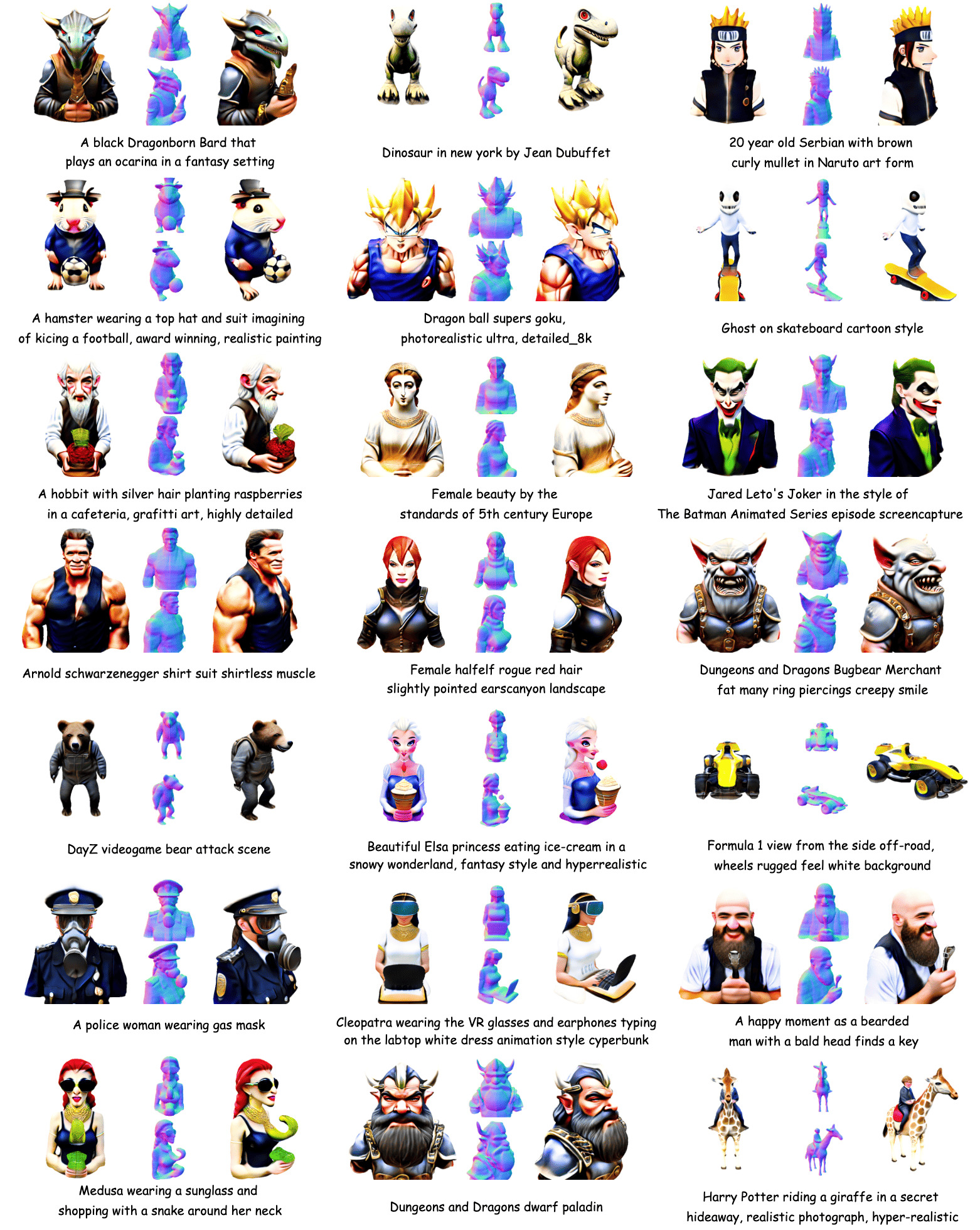}
    \caption{More results of our model trained with expanded corpus.}
    \label{fig:teaser_supp_1}
\end{figure*}

\textbf{Scaling Up Training Corpus}. Since our method is free from the constraints of 3D training data, it can be easily scaled to accommodate more complex and creative text prompts during training. It preserves the SD model's ability to handle creative concepts throughout the 3D adaptation process, generating 3D outputs that faithfully represent the input prompts. As can be seen in the \textbf{+More Text Data} column, when we scale up the training text data from 360K to 1.6M, the CLIP similarity improves by as much as 7\%, leading to more consistent generation for challenging text prompts such as \textit{A lion reading the newspaper} and \textit{A tray of Sushi containing pugs}. This is because our collected text data covers a wider range of creative concepts than the existing 3D datasets~\cite{wu2023omniobject3d,deitke2023objaversexl}, thus providing more sufficient training and improving the generalizability of the model. More visual examples are partially presented in \cref{fig:show_case} and detailed in \cref{fig:teaser_supp_1} and \cref{fig:teaser_supp_2}.

\subsection{Ablation Study}
\label{sec:ablation}

\begin{figure}[!t]
    \centering
    \includegraphics[width=\linewidth]{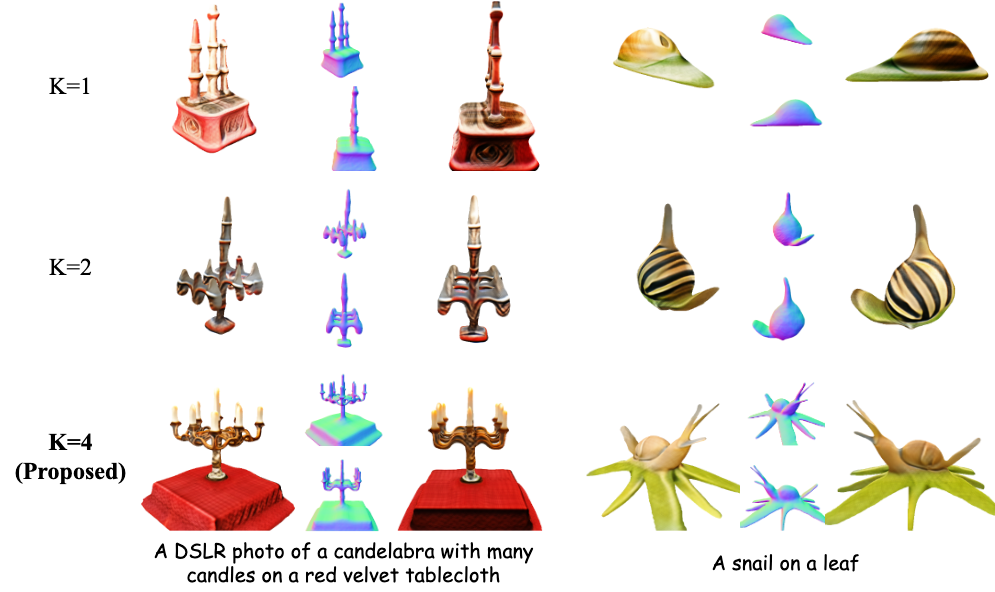}
    \caption{Visualizations on the ablation studies of PRD algorithm. 
    }
    \label{fig:ablation-prd}
\end{figure}

\textbf{The Effectiveness of PRD}. We validate our PRD algorithm by testing a simplified configuration with \textbf{K=1}, which reduces our method to a single-step generation process that is equivalent to a vanilla native 3D generator trained by score distillation~\cite{ma2024scaledreamer}. As shown in \cref{fig:ablation-prd} and quantified in \cref{tab:ablation-prd}, this configuration fails to generate proper 3D structures because it needs to simultaneously handle two complex challenges: adapting SD for 3D generation and performing single-step generation. 
In contrast, our PRD 
~\\~\\~\\
scheme can use alternative step configurations such as \textbf{K=2}, which produces suboptimal yet acceptable results. We found that the configuration of \textbf{K=4} provides the best trade-off between output quality and computational efficiency, which is used as the default setting of PRD.

\begin{table}[!t]
\centering
\begin{tabular}{@{}lll@{}}
\toprule
                 & C.S. $\uparrow$ & R@1 $\uparrow$ \\ \midrule
K=1              & 50.9           & 14.4          \\
K=2              & 62.6         & 22.4         \\
K=4 (Proposed)       &  \textbf{68.2}          & \textbf{32.3}           \\ \bottomrule
\end{tabular}
\caption{Ablation study on the hyper-parameters of PRD.
}
\label{tab:ablation-prd}
\end{table}

\begin{figure}[h]
    \centering
    \includegraphics[width=\linewidth]{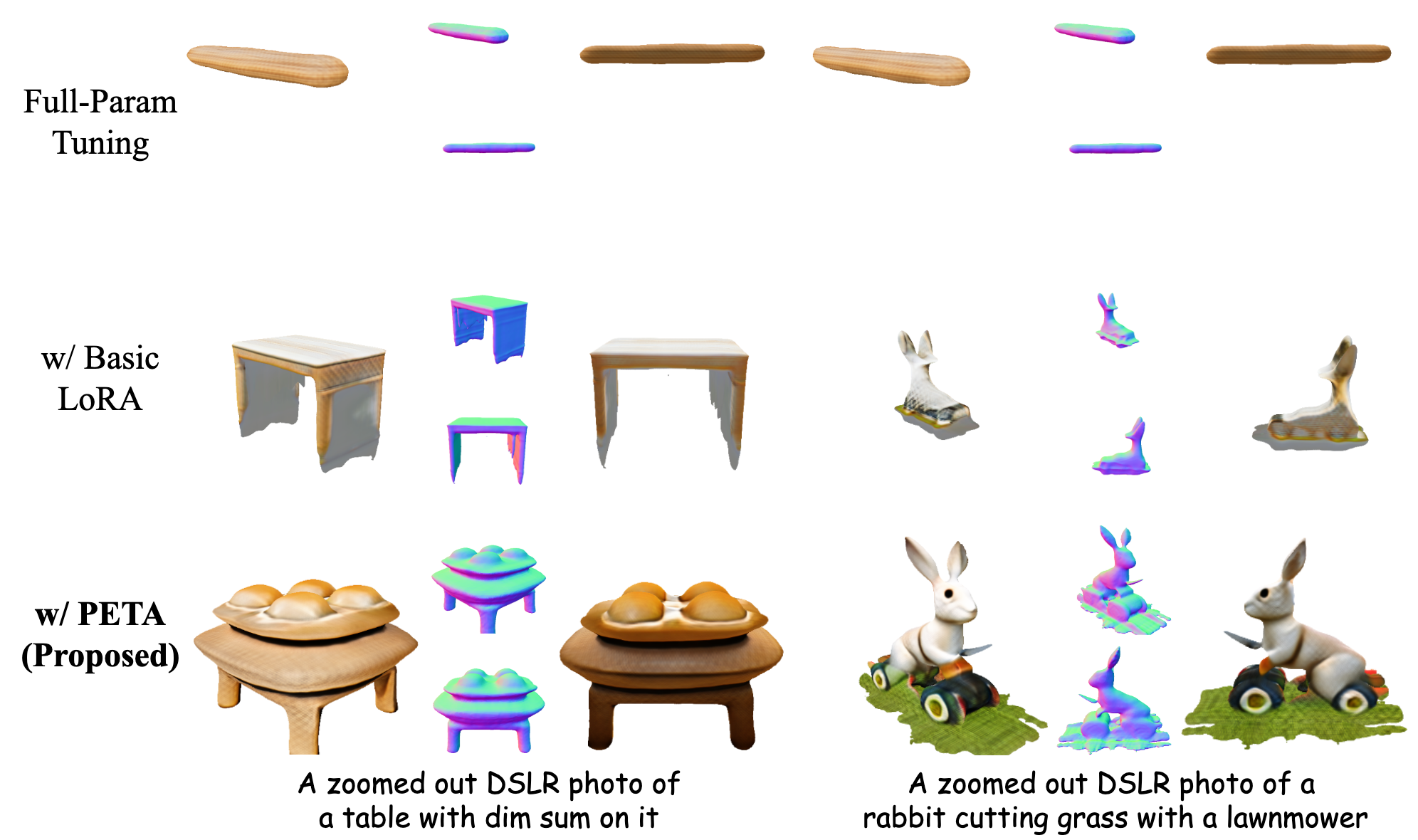}
    \caption{Visualizations on the ablation study of PETA. 
    }
    \label{fig:ablation-peta}
\end{figure}

\textbf{The Effectiveness of PETA}. We first compare our proposed PETA method with conventional full parameter fine-tuning (shown as \textbf{Full Param Tuning} in \cref{fig:ablation-peta} and \cref{tab:ablation-peta}). We see that full parameter fine-tuning exhibits training instability and catastrophic forgetting, resulting in text-inconsistent outputs. We then conduct additional ablation studies using vanilla LoRA tuning, maintaining an equivalent parameter size (22.6M), shown as the configuration of  \textbf{w/ Basic LoRA}. We see that this vanilla adaptation produces degraded geometric structures and textures, and lacks the capability to handle the unique characteristics of each plane. Since different geometry and texture planes (see \cref{fig:peta}) exhibit distinct feature distributions, they require specialized consideration. Our solution (denoted as \textbf{w/ PETA}) considers plane dependency using multiple LoRAs in self-attention blocks, enabling each plane to maintain its unique representation while allowing cross-plane interactions during self-attention computation.   As shown in \cref{fig:ablation-prd} and \cref{tab:ablation-prd}, PETA achieves enhanced 3D generation quality with good text consistency.

\begin{table}[h]
\centering
\begin{tabular}{@{}lll@{}}
\toprule
                 & C.S. $\uparrow$ & R@1 $\uparrow$ \\ \midrule
Full Param Tuning & 35.8      & 0.35         \\
w/ Basic LoRA & 54.2           & 11.1         \\ 

w/ PETA (Proposed)   &  \textbf{68.2}          & \textbf{32.3}          \\  \bottomrule
\end{tabular}
\caption{Ablation study on the effectiveness of PETA. 
}
\label{tab:ablation-peta}
\end{table}

We also perform ablation studies to investigate the impact of multiple multi-view teachers in training, the choice of LoRA rank, our hybrid rendering approach that combines volumetric rendering~\cite{wang2021neus} and mesh rasterization~\cite{wei2023neumanifold} for multi-view distillation in PRD training scheme. The details can be found in \cref{sec:more_ablation}.

\section{Conclusion and Limitation}

In this paper, we presented Progressive Rendering Distillation (PRD), a novel training scheme that adapts Stable Diffusion (SD) for instant text-to-mesh generation without relying on 3D data. We also introduced PETA (Parameter-Efficient Triplane Adaptation), a parameter-efficient method that introduces only $2.5\%$ additional parameters to effectively enable SD for instant text-to-mesh generation. Our model, namely TriplaneTurbo, can produce text-consistent textured meshes in only $1.2$ second. Through comprehensive experiments, we validated the effectiveness of our approach. Our methodology has the potential to be extended to 3D scene generation and image-to-3D tasks. While currently implemented with SD, the PRD approach can also be applied to other pre-trained models like DiT~\cite{esser2024scaling}. We hope our work can inspire new directions in 3D generation to overcome the dependency on 3D data. 

\textbf{Limitations}. One limitation of our method lies in the generation of precise numbers of multiple 3D objects, which may require more sophisticated multi-view teachers, potentially enhanced with layout guidance. Besides, our results for full-body humans might exhibit limited facial and hand details, which can be improved by extending SD adaptation to  more advanced 3D structures than Triplane.

\section*{Acknowledgment}
This work is supported by the InnoHK program.

{
    \newpage
    \small
    \bibliographystyle{ieeenat_fullname}
    \bibliography{main}
}

\newpage
\onecolumn

\customtitle{Supplementary Material to \\
``Progressive Rendering Distillation: Adapting Stable Diffusion for \\ 
Instant Text-to-Mesh Generation without 3D Data''}

\noindent The contents of this supplementary file include:

\begin{itemize}
\item Progressive Rendering Distillation pseudo code (referring to Sec. 3.2 in the main paper).
\item More implementation details (referring to Sec. 4.1 in the main paper).
\item Additional qualitative comparisons (referring to Sec. 4.2 in the main paper).
\item  Additional results with expanded training corpus (referring to Sec. 4.2 in the main paper).
\item Additional ablation experiments (referring to Sec. 4.3 in the main paper).
\end{itemize}

\renewcommand\thesection{\Alph{section}}
\setcounter{section}{0} 

\section{Pesudo Code}
The pseudo-code of our Progressive Rendering Distillation (PRD) training scheme is appended in \cref{alg:progressive rendering distillation}.

\section{More Implementation Details}
\label{sec:more_implementation}


\textbf{Dual rendering}. We integrate DiffMC~\cite{wei2023neumanifold} for mesh rasterization and NeuS~\cite{wang2021neus} for volume rendering to supervise the generation of 3D outputs. Such a dual rendering approach can ensure the training stability: when SDF values are all positive or all negative throughout the 3D space and thus the mesh extraction fails, volume rendering can still guide the training process to optimize the 3D space. Due to memory constraints, volume rendering is limited to low resolution ($128\times128$). We complement this with high-resolution ($512\times512$) mesh rasterization. To handle mesh extraction failures caused by the uniformly distributed SDF signs, we implement the method proposed in \cite{xu2024instantmesh} to artificially enforce the position of the zero-level set in the 3D space. We manually control gradient magnitudes during backpropagation. The gradient of volume rendered multi-views with respect to the texture decoding MLP starts at 1.0 and linearly decreases to 0.01 at the end of training, preventing blurry textures caused by low-resolution volume rendering supervision. The gradient of mesh rasterized multi-views with respect to both the SDF decoding MLP and deformation decoding MLP is fixed at 0.001 throughout training, which stabilizes training and improves generation performance.


\textbf{Training objective}.  With the multi-view teacher~\cite{shi2023MVDream,qiu2023richdreamer,rombach2022high}, we decode the multi-views $\boldsymbol{x}_{\pi}$ to latent $\boldsymbol{z}_{\pi}$, which are diffused by adding Gaussian noise at timestep~$t$~\cite{ho2020denoising}, denoted by $\boldsymbol{z}_{\pi,t}$. We write the diffusion module of the multi-view teacher as $\boldsymbol{z}_{\phi^{2D}}(\boldsymbol{z}_{\pi,t} ; t, \pi, y)$ to represent the process of noise prediction and latent denoising, where $y$ is the text prompt. With ASD \cite{ma2024scaledreamer}, the derivative of the objective  with respect to the 3D generator $\phi^{\mathrm{3D}}$ is:

\begin{equation}
\begin{aligned}
  \nabla_{\phi^{\mathrm{3D}}}\mathcal{L}_{\phi^{\mathrm{2D}}}\left(\boldsymbol{x}_{\pi} ; \pi, y\right)  = 
  \mathbb{E}_{t, \boldsymbol{\epsilon}, \Delta t}\left[
            \omega(t)\left(\boldsymbol{z}_{\phi^{2D}}^{Cls}(
                \boldsymbol{z}_{\pi,t} ; t, \pi, y
            )-\boldsymbol{z}_{\phi^{2D}}(
                \boldsymbol{z}_{\pi, t + \Delta t} ; t + \Delta t, \pi, y
            )
        \right)\frac{
            \partial \boldsymbol{z}_{\pi}
        }{
            \partial \phi^{\mathrm{3D}}
        }\right],
\end{aligned}
\end{equation}
    where $\phi^{2D}$ denotes the teacher model parameters, $t$ is sampled from $\mathcal{U}[T_{\text{Min}}, T_{\text{Max}}]$ with $0<T_{\text{Min}} < T_{\text{Max}} < T = 1000$, and $\text{Cls}$ indicates classifier-free guidance (CFG)~\cite{ho2022classifier}. By introducing a timestep shift $\Delta t$~\cite{ma2024scaledreamer} sampled from a uniform distribution $\mathcal{U}[0, \eta(t - T_{\text{Min}})]$, ASD achieves more effective training of the native 3D generator. We utilize the timestep-dependent weighting factor from DMD~\cite{yin2024one}, as implemented in~\cite{wu2024one,sun2024pixel}. We let

\begin{equation}
\begin{aligned}
    \omega(t) = \frac{1}{\mathrm{NoGrad}(
            \mathrm{Mean}(
                \boldsymbol{z}_{\pi} - 
                \boldsymbol{z}_{\phi^{2D}}^{Cls}(\boldsymbol{z}_{\pi,t} ; t, \pi, y) 
            )
        )+ 
        \delta
    },
\end{aligned}
\end{equation}
where $\mathrm{NoGrad}$ detaches gradients for loss back-propagation, and $\mathrm{Mean}$ applies $L_1$-norm across height, width, channel dimensions and all rendered views. Unlike~\cite{wu2024one,sun2024pixel,yin2024one}, we add constant $\delta=0.1$ to the denominator, which stabilizes training and improves generation performance. We apply this objective function to supervise 3D outputs using three teacher models (SD, MV, RD) and two rendering pipelines (volume rendering and mesh rasterization). Regarding the sampling range of timestep $t$, $T{_\text{Max}}=980$ throughout training, while $T_{\text{Min}}$ starts at 500 and linearly decreases to 20. Teacher-specific hyperparameters vary: RD uses CFG=20 and $\eta=0.1$; MV uses CFG decreasing from 20 to 10 and $\eta=0$; SD uses CFG=5 and $\eta=0$. Setting $\eta=0$ for multi-view teachers that supervise RGB renderings aligns with the findings in PiSA-SR~\cite{sun2024pixel}.  Additionally, we incorporate regularization terms during training, such as sparsity loss~\cite{poole2022dreamfusion} and eikonal loss~\cite{yariv2021volume}. We linearly reduce the sparsity and eikonal loss weights from 1 to 0 throughout the training process.

\begin{figure*}[!t]
    \centering
    \includegraphics[width=0.95\linewidth]{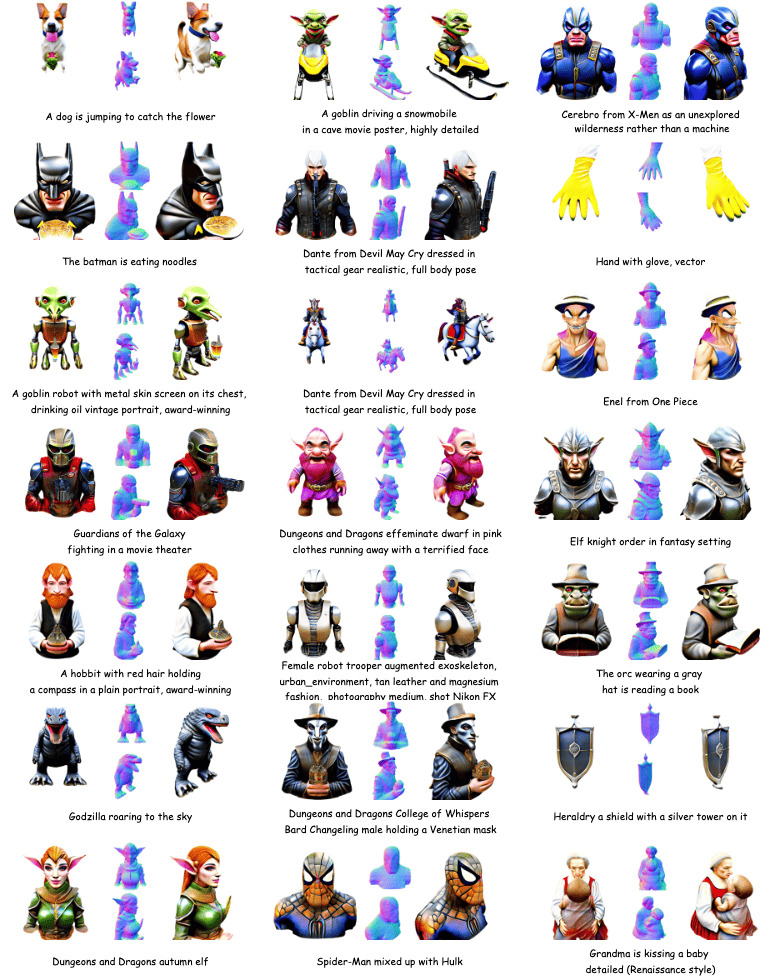}
    \caption{More results of our model trained with expanded corpus.
    }
    \label{fig:teaser_supp_2}
\end{figure*}

\textbf{Noise schedule}. The PRD training incorporates progressive noise addition to the denoised latents (see \cref{alg:step:add_noise} in \cref{alg:progressive rendering distillation}). Being adapted from SD~\cite{rombach2022high}, our native 3D generator follows the DDPM~\cite{ho2020denoising} noise schedule in training. During inference, we employ DDIM~\cite{song2020denoising}.

\begin{algorithm}[!t]
    \caption{Progressive Rendering Distillation (PRD)}
    \label{alg:progressive rendering distillation}


    \KwIn{
    SD-based native 3D generator with $\boldsymbol{z}_{\phi^{\mathrm{3D}}}$ and $\emph{D}_{\phi^{\mathrm{3D}}}$;
    score distillation objective $\mathcal{L}_{\phi^{2D}}$ parameterized by multi-view diffusion model $\phi^{2D}$;
    prompt corpus $\mathbb{S}_y$; number of rendered views $N$; number of steps $K$
    }

    Initialize optimizer $\mathrm{Opt}$ for $\boldsymbol{z}_{\phi^{\mathrm{3D}}}$ and $\emph{D}_{\phi^{\mathrm{3D}}}$

    Define fixed timesteps $T = t_1 > t_2 > \cdots t_K > 0$

    \While{not converged}{        

        Sample text prompt $y \in \mathbb{S}_y$

        Sample $\hat{\boldsymbol{z}_0} \sim \mathcal{N}(0, \boldsymbol{I})$

        \For{$t \gets t_1$ \KwTo $t_K$}{        

            Sample $\epsilon \sim \mathcal{N}(0, \boldsymbol{I})$
            
            $\boldsymbol{z}_{t} \gets \alpha_{t} \hat{\boldsymbol{z}_0} + \sigma_{t} \epsilon$ \label{alg:step:add_noise}

            $\hat{\boldsymbol{z}_0} \gets \boldsymbol{z}_{\phi^{\mathrm{3D}}}(
                \boldsymbol{z}_t; t, y
                )
            $

            $ \hat{\theta} \gets \emph{D}_{\phi^{\mathrm{3D}}}(
                \hat{\boldsymbol{z}_0}
            )
            $

            Sample $K$ camera poses $\pi_1, \dots, \pi_N$



            \For{$i \gets 1$ \KwTo $N$}{        
                $\boldsymbol{x}_{\pi_i} \gets g(\hat{\theta}, \pi_i)
                $
            }
            $\boldsymbol{L} \gets \mathcal{L}_{\phi^{2D}}(\boldsymbol{x}_{\pi_1}, \dots, \boldsymbol{x}_{\pi_N}; \pi_1, \dots, \pi_N, y)$

            Save $\frac{1}{K}\nabla_{\phi^{\mathrm{3D}}}\boldsymbol{L}$ in $\mathrm{Opt}$

        }
        Update $\boldsymbol{z}_{\phi^{\mathrm{3D}}}$ and $\emph{D}_{\phi^{\mathrm{3D}}}$ with gradient saved in $\mathrm{Opt}$
    }

    \Return $\boldsymbol{z}_{\phi^{\mathrm{3D}}}$ and $\emph{D}_{\phi^{\mathrm{3D}}}$ 
 
\end{algorithm}

\begin{figure}[!t]
    \centering
    \includegraphics[width=0.6\linewidth]{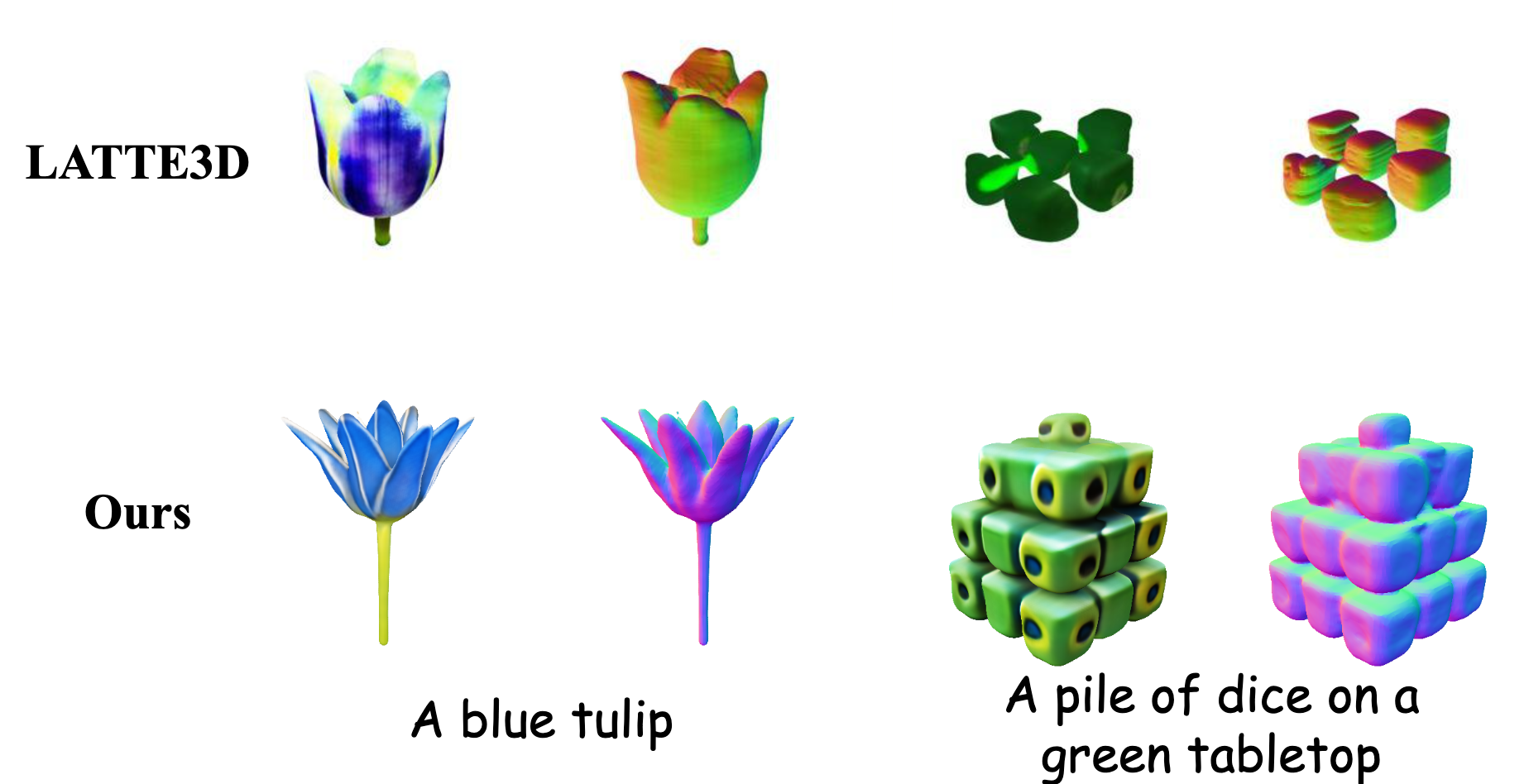}
    \caption{Qualitative comparison with LATTE3D~\cite{xie2024latte3d}.
    }
    \label{fig:comparison_latte3D}
\end{figure}

\begin{figure}[!t]
\begin{minipage}[t]{0.5\linewidth}
\centering
\includegraphics[width=\linewidth]{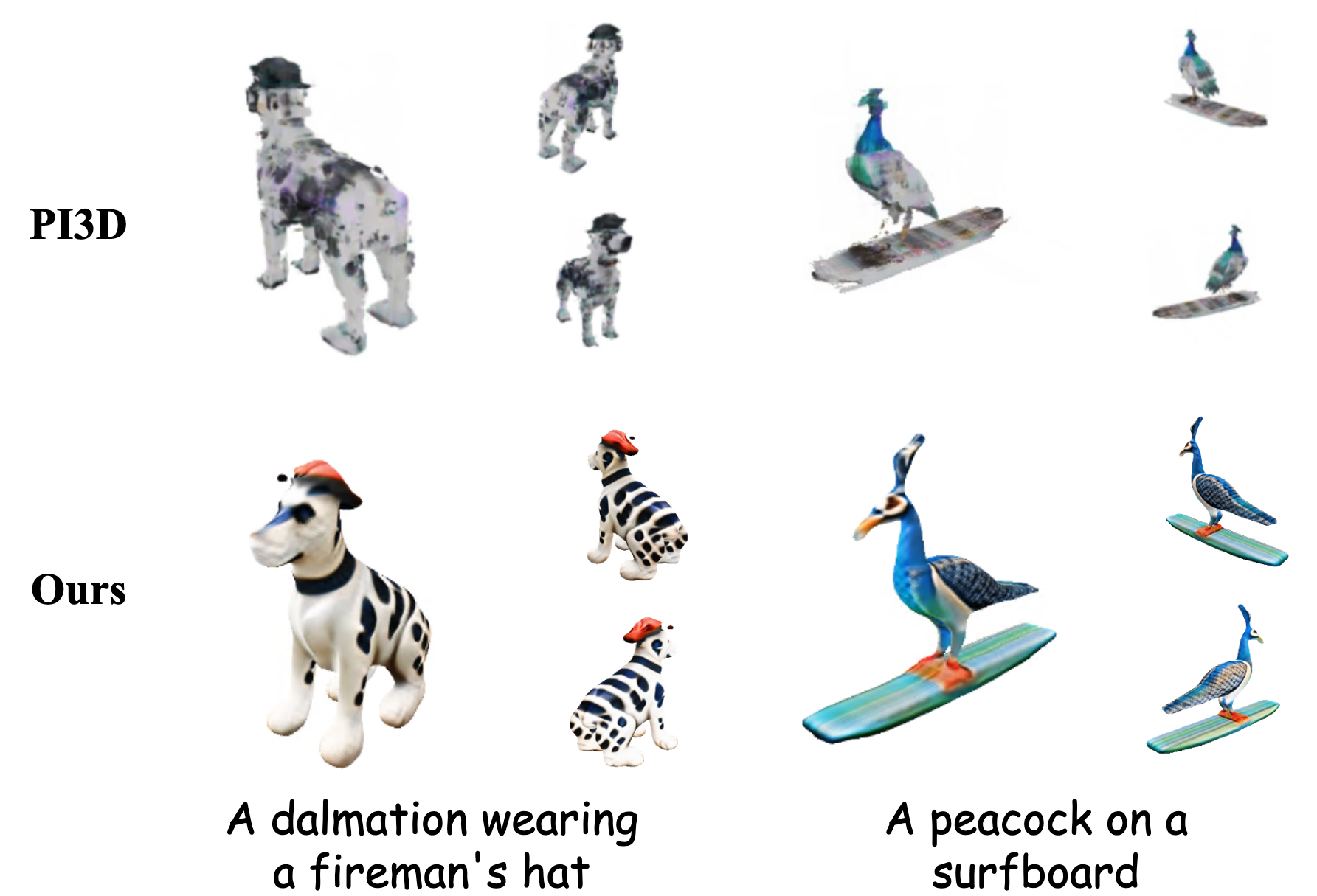}
\caption{Qualitative comparison with PI3D~\cite{liu2024pi3d}.}
\label{fig:comparison_pi3d}
\label{}
\end{minipage}%
\begin{minipage}[t]{0.5\linewidth}
\centering
\includegraphics[width=\linewidth]{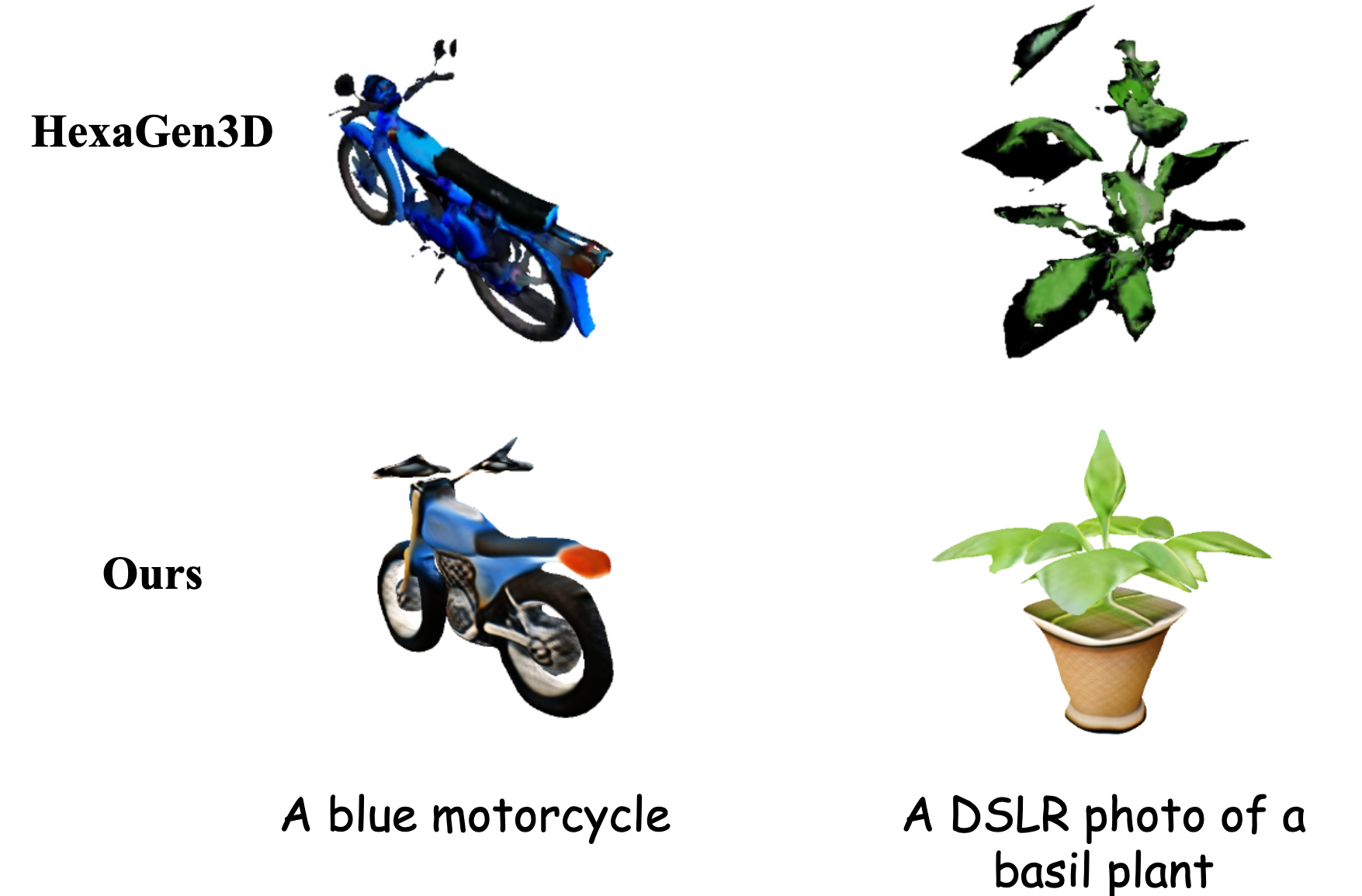}
\caption{Qualitative comparison with HexaGen3D~\cite{mercier2024hexagen3d}.}
\label{fig:comparison_hexagen3D}
\end{minipage}
\end{figure}

\section{More Qualitative Comparison Results}

\textbf{Comparison with methods adapting SD as native 3D generators}. Since the codes or trained models of current SD-based native 3D generators~\cite{mercier2024hexagen3d,liu2024pi3d} are not publicly available, we conduct our comparisons by using their visual results presented in the original publications. The qualitative comparisons with PI3D~\cite{liu2024pi3d} and HexaGen3D~\cite{mercier2024hexagen3d} are presented in~\cref{fig:comparison_pi3d} and~\cref{fig:comparison_hexagen3D}, respectively. As both the two compared methods employ data-driven training, they inherit pose inconsistencies existed in the 3D training datasets~\cite{deitke2023objaversexl}, leading to the issue of occasional pose misalignment. This can be clearly observed from PI3D's result of \textit{`A dalmatian wearing a fireman's hat'} shown in~\cref{fig:comparison_pi3d}, where the dog is oriented sideways. The comparison results demonstrate our method's superior visual fidelity with the input prompts. These improvements are attributed to our proposed Progressive Rendering Distillation (PRD) scheme, which utilizes multi-view teachers in training without requiring 3D training data. 

\textbf{Comparison with native 3D generators trained with score distillation}. We further compare our approach with existing methods that employ score distillation for native 3D generator training. Specifically, we compare against the current state-of-the-art method, LATTE3D~\cite{xie2024latte3d}. Since the code or model of LATTE3D is unavailable, we conduct qualitative comparisons using their published results. The visual comparisons are presented in~\cref{fig:comparison_latte3D}. It can be seen that our method demonstrates notable improvements in both texture fidelity and geometric accuracy. For example, in \textit{`A blue tulip'}, our model captures more natural flower textures, while in \textit{`A pile of dice on a green tabletop'}, our model achieves more precise geometric structures. These improvements can be attributed to our strategic adaptation of SD as the backbone architecture, which allows our model to leverage its powerful generative capabilities.

\section{Expanding Training Corpus}
\label{sec:more_corpus}

Since our proposed training scheme does not require 3D ground truth data, it can be easily up-scaled to a large amount of text prompts. We collect a total number of 1.7 million text prompts from HuggingFace that were used to generate images by DALL-E and Midjourney. This corpus has more unnatural prompts than the Objaverse~\cite{deitke2023objaversexl}, and it is more challenging. To the best of our knowledge, our work is the first that can process more than 1 million training data. Our model, trained on this expanded dataset, achieves enhanced visual quality, as demonstrated in Fig. 1 in the main paper and \cref{fig:teaser_supp_1}, \cref{fig:teaser_supp_2} in this supplementary file.

\section{More Ablation Studies}
\label{sec:more_ablation}

\textbf{The necessity of multiple teachers}. We employ SD \cite{rombach2022high}, MV \cite{shi2023MVDream} and RD \cite{qiu2023richdreamer} as teachers for multi-view supervision of RGB, normal and depth maps. Here we perform ablation studies by systematically removing individual components. 

\begin{figure}[!t]
    \centering
    \includegraphics[width=0.68\linewidth]{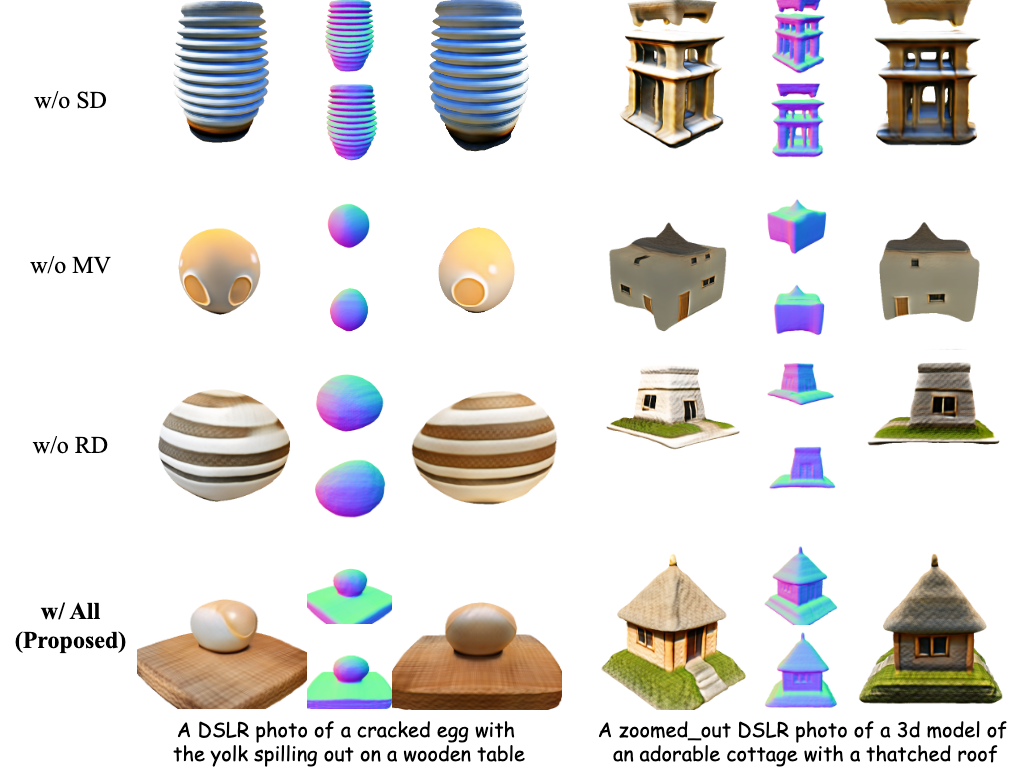}
    \caption{Visualizations for the ablation study on jointly using SD, MV and RD as multi-view teachers. 
    }
    \label{fig:ablation_teachers}
\end{figure}

First, as visualized by \textbf{w/o SD} in \cref{fig:ablation_teachers}, when SD is removed, leaving only MV and RD as teachers, the model can collapse to generate results inconsistent with text prompts. For example, given the prompt \textit{`A DSLR photo of a cracked egg with the yolk spilling out on a wooden table'}, the model collapses to generating a stack of discs. This occurs because training for multi-view generation may impair MV and SD's text understanding capabilities, resulting in outputs that diverge from the specified text descriptions. SD can prevent from training collapse and improve the generation stability.  
Second, the importance of MV is demonstrated by the visualizations of \textbf{w/o MV} in \cref{fig:ablation_teachers}. Without multi-view RGB supervision, the generated results tend to show repetitive and redundant contents across different viewpoints. For instance, multiple \textit{`egg yolks'} might appear in the results of  \textit{`A DSLR photo of a cracked egg with the yolk spilling out on a wooden table'}. 
Finally, the importance of RD is validated by the visualizations of \textbf{w/o RD}. We can see that adding normal and depth constraints enhances text consistency in the outputs, such as the generated \textit{`wooden table'} in the results of \textit{`A DSLR photo of a cracked egg with the yolk spilling out on a wooden table'}. 
Overall, the combination of SD, MV, and RD as teachers achieves the best results, as validated by the visualization of \textbf{w/ All} and the metrics shown in \cref{tab:ablation-teachers}.

\begin{table}[!t]
\centering
\begin{tabular}{@{}lll@{}}
\toprule
                 & C.S. $\uparrow$ & R@1 $\uparrow$ \\ \midrule
w/o SD           &   63.0          & 20.1         \\
w/o MV           &  67.4           & 25.9         \\
w/o RD           & 41.5           & 11.4         \\ \midrule
w/ All (Proposed)       &  \textbf{68.2}          & \textbf{32.3}           \\ \bottomrule
\end{tabular}
\caption{Ablation study on jointly using SD, MV and RD as multi-view teachers.
}
\label{tab:ablation-teachers}
\end{table}

\begin{figure}[!h]
    \centering
    \includegraphics[width=0.7\linewidth]{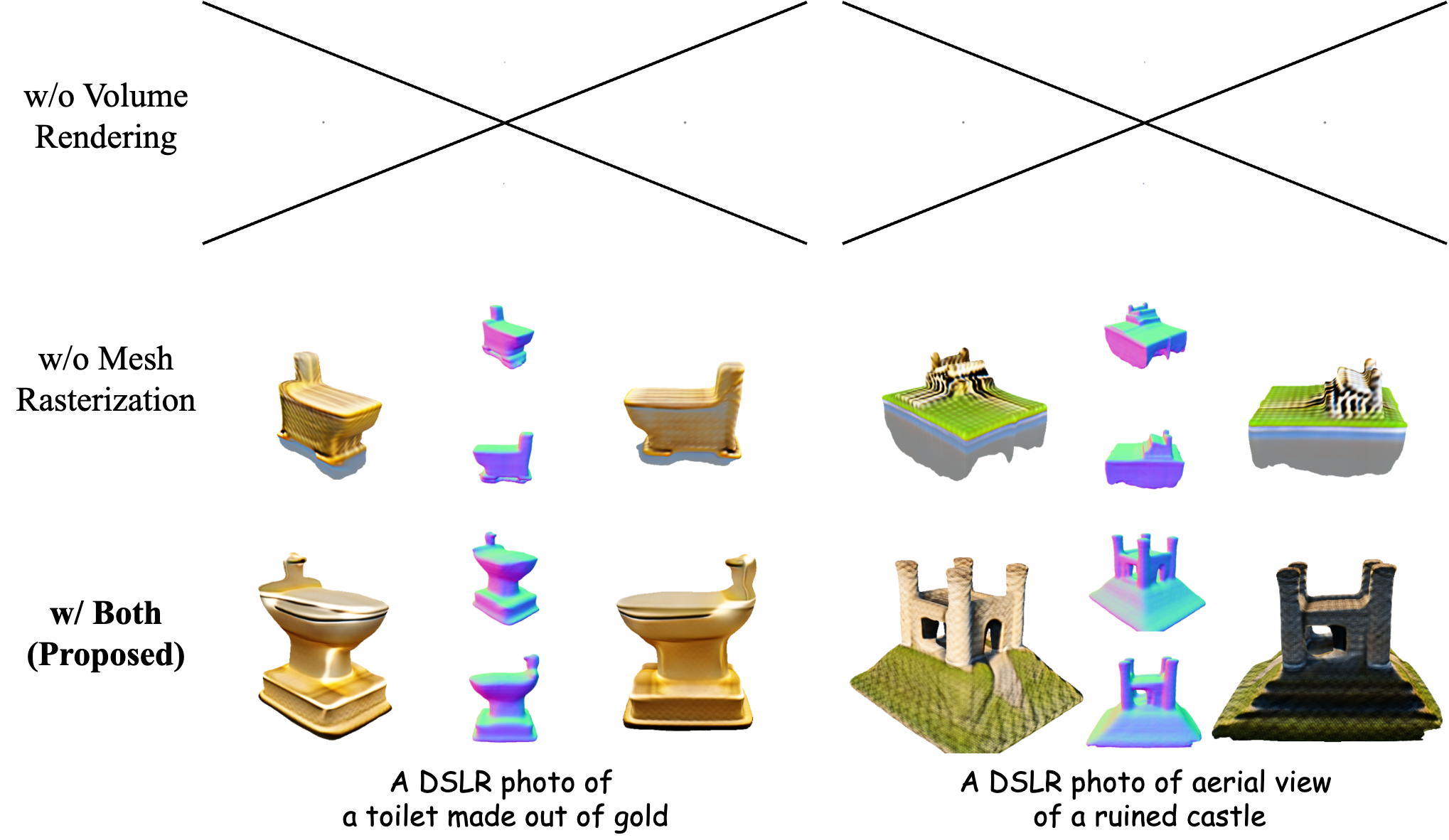}
    \caption{Ablation study on dual rendering. The cross mark means the model fails to generate mesh due to training instability.
    }
    \label{fig:ablation_renderers}
\end{figure}

\textbf{The necessity of dual rendering}. We use a dual rendering framework that integrates mesh rasterization~\cite{kim2023neuralfield} and volume rendering~\cite{yariv2021volume} for 3D output supervision, as detailed in~\cref{sec:more_implementation}. The effectiveness of this dual approach is demonstrated through quantitative and qualitative evaluations in~\cref{tab:ablation-renderers} and~\cref{fig:ablation_renderers}, respectively. Without volume rendering, relying solely on mesh rasterization leads to training collapse and invalid mesh extraction. The results labeled as \textbf{w/o Volume Rendering} in~\cref{fig:ablation_renderers} demonstrate that training converges to a state where the SDF's zero-level set vanishes, resulting in mesh extraction failure and empty space. 
Conversely, using only volume rendering, which is constrained to low-resolution training, fails to produce high-quality mesh geometry, leading to rough and coarse textural details, as shown by \textbf{w/o Mesh Rasterization} in~\cref{fig:ablation_renderers}. For example, it fails to produce the shining gold texture for the prompt \textit{`A DSLR photo of a toilet made out of gold'}. Moreover, without direct mesh supervision, volume rendering-based methods may produce geometrically invalid structures. This limitation is evident in the result of \textit{'A DSLR photo of aerial view of a ruined castle'}, where the extracted meshes exhibit incorrect structural features and poor textures, manifesting as gray regions in parts of the mesh. As shown by \textbf{w/ Both} in~\cref{fig:ablation_renderers} and supported by the superior metrics in~\cref{tab:ablation-renderers}, our dual rendering approach enables stable training while producing meshes with detailed textures and well-defined geometric structures.

\begin{figure}[!t]
    \centering
    \includegraphics[width=0.7\linewidth]{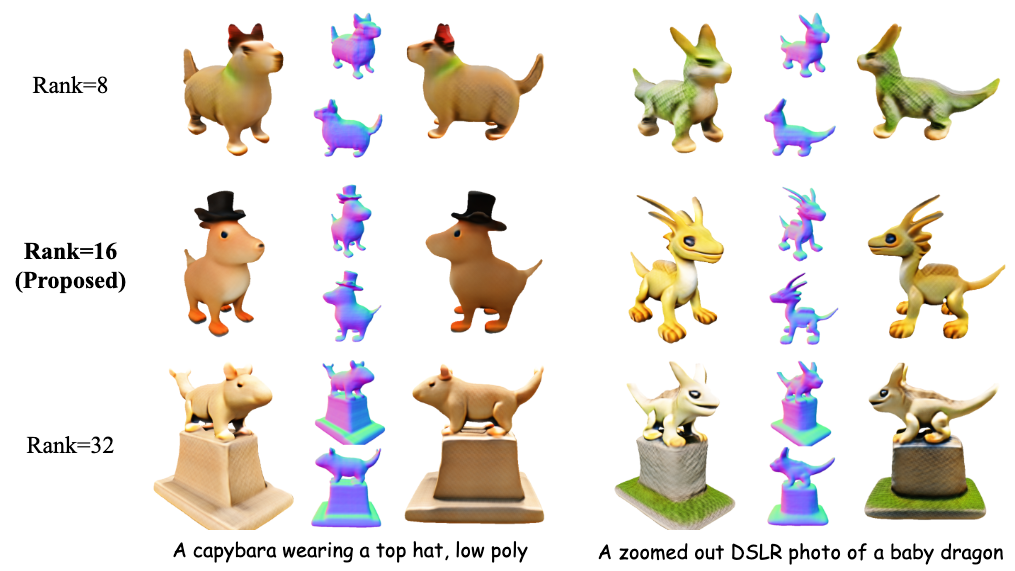}
    \caption{Visualization for the ablation study on the LoRA rank in PETA.
    }
    \label{fig:ablation_lora_rank}
\end{figure}

\vspace{+2mm}
\textbf{The choice of LoRA rank}. We demonstrate the significance of using a LoRA rank of 16 in our Parameter-Efficient Triplane Adaption (PETA). With a lower rank of 8, shown as \textbf{Rank=8} in~\cref{fig:ablation_lora_rank}, the model exhibits insufficient learning capacity, as evidenced by its failure to generate the top hat structure for the prompt \textit{`A capybara wearing a top hat, low poly'}. However, setting a higher rank, such as 32, can also lead to unreasonable geometric outputs. As shown in \textbf{Rank=32} in~\cref{fig:ablation_lora_rank}, some unwanted platform structures appear at the bottom of results of \textit{`A capybara wearing a top hat, low poly'} and \textit{`A zoomed out DSLR photo of a baby dragon'}. Such artifacts stem from the inherent generation biases in both MV and SD, as their training dataset~\cite{deitke2023objaversexl} contains numerous examples where objects rest on square platforms. As a result, the multi-view teachers are fitted to generate outputs with similar structures. 
When the LoRA rank is set too high, the native 3D generator tends to learn and reproduce the biases from the teachers during the distillation. Setting the rank to a balanced value of 16 enables the model to generate text-aligned 3D results while avoiding the incorporation of undesirable biases into the 3D generation model. Denoted as \textbf{Rank=16}, both qualitative results in~\cref{fig:ablation_lora_rank} and quantitative results in~\cref{fig:ablation_lora_rank} show that a rank of 16 yields the best performance.

\begin{minipage}{\textwidth}
\begin{minipage}[!t]{0.48\textwidth}
\makeatletter\def\@captype{table}
\centering
\scalebox{0.9}{
\begin{tabular}{@{}lll@{}}
    \toprule
                 & C.S. $\uparrow$ & R@1 $\uparrow$ \\ \midrule
w/o Volume Rendering           &    25.1          & 0.01         \\
w/o Mesh Rasterization           &  67.4           & 25.9         \\ \midrule
Joint (Proposed)       &  \textbf{68.2}          & \textbf{32.3}           \\ \bottomrule
\end{tabular}
}
\caption{Ablation study on the dual renders.}
\label{tab:ablation-renderers}
\end{minipage}
\begin{minipage}[!t]{0.48\textwidth}
\makeatletter\def\@captype{table}
\centering
\begin{tabular}{@{}lll@{}}
    \toprule
                 & C.S. $\uparrow$ & R@1 $\uparrow$ \\ \midrule
rank=8           &    62.9          & 15.6         \\
rank=16 (Proposed)       &  \textbf{68.2}          & \textbf{32.3}   \\        
rank=32           &  66.2           & 26.6         \\   \bottomrule
\end{tabular}
\caption{Ablation study on the LoRA rank in PETA.}
\label{tab:ablation-rank}
\end{minipage}
\end{minipage}

\end{document}